\documentclass[aps,prd,amsmath,amssymb,showpacs]{revtex4}
\usepackage{graphics,epsfig,epsf,psfrag}
\usepackage{bm}
\usepackage{color}
\usepackage{amsmath}
\usepackage{amssymb}
\usepackage{amsfonts}
\usepackage{keyval,graphicx}
\usepackage{textcomp,wasysym}
\newcommand{\jpsi}{J/\psi}

\newcommand{\intq}{\int\!\frac{\mathrm d^3q}{(2\pi)^3}}
\newcommand{\kev}{~\mathrm{keV}}
\newcommand{\mev}{~\mathrm{MeV}}
\newcommand{\gev}{~\mathrm{GeV}}
\newcommand{\jc}{{\cal J}}

\begin{document}
%=====================================================================================================
\title{Properties of open and hidden charm mesons in light quark matter}
%=====================================================================================================
\author{Martin Cleven$^{1}$\footnote{{\it E-mail address:} cleven@fqa.ub.edu},
        Volodymyr K. Magas$^{1}$\footnote{{\it E-mail address:} vladimir@fqa.ub.edu},
        and Angels Ramos$^{1}$\footnote{{\it E-mail address:} ramos@fqa.ub.edu}
        }
        
\affiliation{$^1$ Departament de Fisica Quantica i Astrofisica and Institut de Ciencies del Cosmos\\
        Universitat de Barcelona, 08028-Barcelona, Spain}
%=====================================================================================================
\begin{abstract}
In this work we study the implications of light-quark pionic matter at finite temperatures on the properties of open and hidden charm mesons. 
The meson-meson interactions are described by means of a chiral unitary approach accounting for coupled channels effects. 
The in-medium Lippmann-Schwinger Equations, which consider the change in self-energy that the mesons acquire from interacting with the surrounding pionic matter, are solved self-consistently, and the spectral functions of the mesons in the hot pion bath are obtained. It is observed that the charmed mesons develop a quite substantial pion-induced width, being of several tens of MeV at a temperature of 150 MeV. The $\jpsi$ meson stays narrow, but its pionic width at 150 MeV, found to be around 0.1 MeV, is already larger that its vacuum width. 
\end{abstract}
%=====================================================================================================

\maketitle

%==================================================
\section{Introduction}
%==================================================
%
With a new generation of heavy-ion-collision experiments  operational or under construction --LHC, FAIR or NICA-- there is a demand for theoretical predictions of hadronic properties at temperatures and densities far from standard nuclear physics scenarios. 
Among the various subjects, the study of the $\jpsi$ stands out as a candidate to signal deconfinement, according to the Matsui and Satz prediction \cite{MatsuiSatz}. 
A quark-gluon plasma (QGP) produced in the collision would screen the $c\bar c$ interaction or ionize the charmonium state. 
Either way, this would lead to a suppression of events.

However, although in previous experiments such a drop was actually seen \cite{NA50}, it still remains unclear whether this is indeed related to the formation of a QGP. 
The inelastic interactions of the $\jpsi$ with the surrounding hadronic medium, accounted for in the  'co-mover' models, offer alternative mechanisms to explain the drop in the $\jpsi$ production (see \cite{comover} and Refs. therein) that either do not involve the transition to QGP state or, at least, reduce the number of $\jpsi$ during the evolution in the hadronic phase of reaction, and thus have to be taken into account.  

The properties of open and hidden charm mesons are especially interesting in view of the alternative scenario for charm production in ultrarelativistic heavy ion collisions based on the statistical production model. Such scenario was proposed for the first time in the work of Gorenstein and Gazdzicki \cite{Gazdzicki:1999rk} (for more recent reviews see \cite{SHM,Stachel_talk}). According to this model, particles with charm are produced in exactly the same way as those built from the light quarks: at the moment of chemical freeze out all the different hadron species are produced with corresponding equilibrium Bose or Fermi distributions depending on local temperature, corresponding chemical potentials (baryonic, strange, charm) and reaction volume. For the further (post freeze out) evolution of the spectra, the resonance decays have to be taken into account. 
To eliminate the volume dependence the ratios of different hadrons are usually studied. 
One particular property of the statistical production model is that the particle ratios do not depend on the initial collision energy, but only on the freeze out conditions.  In such a scenario, for example, the ratio of hidden to open charm mesons, $\jpsi/D$, will depend, with a good accuracy, only on the chemical freeze out temperature, which actually saturates around $T=160$ MeV for the central heavy ion collisions already at RHIC. Of course, for particles with charm quarks the statistical production becomes applicable only at sufficiently high collision energies, when the local chemical equilibrium for the charm charge is achieved. And the recent results from ALICE seem to support the idea that charm can reach the equilibrium stage in heavy ion collisions at LHC energies \cite{Stachel_talk}. 

The applicability of statistical production model for particles with charm opens a very interesting possibility of studying the production of all known charmed resonances in ultra-relativistic heavy ion collisions, including exotic resonances like tetraquarks,  as the $X(3872)$ state - the first tetraquark discovered at Belle detector \cite{x3872}, and pentaquarks, as the $P_c(4380)$ and $P_c(4450)$ resonances recently observed in  $\Lambda_b \rightarrow \jpsi K- p$ decay \cite{P_c}.  All such resonances will be produced in statistically determined amounts at the chemical freeze out hypersurface, and their probability to survive during the interacting hadronic phase of the collision and finally reach the detector will depend on their interaction with other hadrons at finite temperature. Therefore, the modifications induced by a hot light quark matter on these interactions should be addressed to properly understand the charm production in heavy ion collisions. Thus, we have a possibility to test different models of the charmed hadrons, in particular the chiral model with unitarization imposed in coupled channels, which generates resonances having a molecular like structure. For example, the $X(3872)$ resonance has been explained as $D\bar{D}^*+c.c.$  molecule \cite{mol_x3872}, and thus we can aim to calculate its spectral function at finite temperature once we know the finite temperature behaviour of the charmed mesons. 

Obtaining the properties of the $D$ and $D^*$ mesons in hot pionic matter, which we consider to be the first level approximation of the matter generated in ultrarelativistic heavy ion collision, will be one of the main topics of the present study.  While interesting in itself, the study of these mesons is also closely related to that of the $\jpsi$ through coupled-channel effects. Therefore, the second focus will be on studying the properties of the hidden charm $\jpsi$ meson under the same conditions. Such a combined effort will provide a comprehensive understanding of how the properties of open and hidden charm mesons are modified when interacting with pionic matter at non-vanishing temperatures.

%---------------------------
% Previous Studies
%---------------------------

Previous studies about  $\jpsi$-hadron interaction at finite temperature broadly fall into two categories: interactions based on chiral Lagrangians~\cite{Haglin:2000ar,Bourque:2008ta,Blaschke:2008mu,Blaschke:2012zza} and quark model calculations~\cite{Zhou:2012vv,Maiani:2004py,Maiani:2004qj,Bourque:2008es}. Zhao and Rapp~\cite{Zhao:2010nk} use thermal lattice QCD to constrain the in-medium charmonium properties in order to calculate spectral functions of the $\jpsi$.
A phenomenological study of the $D$, $D^*$ and $\jpsi$ mesons in a hot pion medium based on phenomenologically built resonant amplitudes  and employing ad-hoc factors to account for thermal effects is also available \cite{Fuchs:2004fh}.
The pion induced width of the $D$-meson has also been calculated by He et al. in ~\cite{He:2011yi} using the resonance-based phenomenological approach.
While some of those works have gone a long way in explaining the reduction of $\jpsi$ production in previous experiments, the calculations typically lack some of the state-of-the-art techniques that have been developed in the subsequent years. 
With this in mind and the need for updated predictions due to the next generation of experiments we will improve the calculations compared to previous studies in two major ways.
First, we will use unitarized coupled channel amplitudes. This has a sizable impact on the dissociation cross sections for the  $\jpsi$, as we will see.
Secondly, we will use the Imaginary Time Formalism (ITF) to rigorously introduce temperature and many-body effects. We focus on the s-wave interactions of the heavy-mesons with the surrounding mesons, which give the most relevant contributions to the self-energy of these mesons at rest. 
A field-theoretical approach studying the one-loop p-wave self-energy of the $D$ and $D^*$ mesons in a hot hadronic medium can be found in~\cite{Ghosh:2013xea}. However, these contributions will not be considered in this work, as we are focussion on the properties of these mesons at rest. 
In Ref.~\cite{Abreu:2011ic} Abreu et al. calculate the drag and diffusion coefficients of charmed mesons in unitarized effective theory, a task we will leave for a future application of our framework.

This work is structured as follows: In Sec.~\ref{Sec:Framework} we will give an overview of the $SU(4)$ model developed by Gamermann et al.~\cite{Gamermann:2006nm,Gamermann:2007fi} 
followed by the modifications using ITF to introduce finite temperatures and densities. 
Results will be presented in Sec.~\ref{Sec:results}. We will finish with a short outlook and a brief summary. 

%===============================================================
\section{Theoretical Framework}\label{Sec:Framework}
%===============================================================

%===============================================================
\subsection{Unitarized amplitudes in the vacuum}\label{Sec:su4}
%===============================================================
We will begin by laying out the foundations to calculate meson-meson scattering in a chiral $SU(4)$ model. 
This model was succesfully applied in various works -- see $e.g.$ Refs.~\cite{Gamermann:2006nm,Gamermann:2007fi}. 
We will only focus on the essential points here and refer to these references for further reading. 

The pseudoscalar and vector fields are collected in the $SU(4)$ 15-plets $\Phi$ and $\mathcal V_\mu$, respectively, which are defined as

{\begin{footnotesize}
\begin{eqnarray}
\Phi= \left( \begin{array}{cccc}
\frac{1}{\sqrt{2}}\pi^0+\frac{1}{\sqrt{3}}\eta & \pi^+ & K^+ & \bar D ^0\\
\pi^- & -\frac{1}{\sqrt{2}}\pi^0+\frac{1}{\sqrt{3}}\eta &  K^0 & D^-\\
K^- & \bar{K^0} & \sqrt{\frac{2}{3}}\eta  & D_s^- \\
D^0 & D^+ & D_s^+ & \eta_c
\end{array} \right), \quad
\mathcal V_\mu= \left( \begin{array}{cccc}
\frac{1}{\sqrt{2}}\rho^0_\mu+\frac{1}{\sqrt{2}}\omega_\mu & \rho^+_\mu & K_\mu^{*+} & \bar D_\mu ^{*0}\\
\rho^-_\mu & -\frac{1}{\sqrt{2}}\rho_\mu^0+\frac{1}{\sqrt{2}}\omega_\mu &  K_\mu^{*0} & D_\mu^{*-}\\
K^{*-} & \bar{K}_\mu^{0*} & \phi_\mu & D_{s\mu}^{*-} \\
D_\mu^{*0} & D_\mu^{*+} & D_{s\mu}^{*+} & \psi_\mu\end{array} \right).
\end{eqnarray}
\end{footnotesize}
}

For each of the two one can construct a vector current
\begin{eqnarray}
  J_\mu = (\partial_\mu\Phi)\Phi - \Phi(\partial_\mu\Phi) ,\qquad
  \jc_\mu = (\partial_\mu \mathcal V_\nu)\mathcal V^\nu - \mathcal V_\nu(\partial_\mu \mathcal V^\nu). 
\end{eqnarray}
Connecting the two we can obtain the interaction Lagrangian
\begin{equation}\label{Eq:lag}
 \mathcal{L_\mathrm{PPPP}} = \frac{1}{12f^2}\left<J^{\mu}J_{\mu}+\Phi^4M\right> \qquad
 \mathcal{L_\mathrm{VPVP}} = -\frac{1}{4f^2}\left<J^{\mu}  \jc_\mu\right>  \ ,
\end{equation}
where $\left<... \right>$ denotes the trace over flavor indices.
In the usual case of SU(3) the parameter $f$ is the pion decay constant, $f=f_\pi$. However, since we are to a large extend interested in the interactions of heavy mesons we need to account for the fact the $D$-meson decay constant is larger, $f_D=165\mev$.
This is done by replacing the $f^2$ factor in the vertex with $\sqrt{f_\pi}$ for each light meson leg and $\sqrt{f_D}$ for each heavy one. 
Finally, the mass term that gives rise to the $\Phi^4$ interaction for pseudoscalar mesons in $SU(4)$ is given by
\begin{equation}
 M = \left( \begin{array}{cccc} m_\pi^2 &0&0&0 \\ 0 & m_\pi^2 &0&0 \\ 0&0&2m_K^2-m_\pi^2 &0 \\ 0&0&0& 2m_D^2-m_\pi^2 \end{array} \right).
\end{equation}

Since $SU(4)$ is not an exact symmetry in nature one needs to break it in an appropriate way. 
This is done with factors that correspond to the leading allowed $t$-channel exchange -- $\gamma = (m_L/m_H)^2$ for charmed mesons and $\psi = -1/3+4/3(m_L/m'_H)^2$ for charmonia. 
We will use $m_L=800\mev$, $m_H=2050\mev$ and $m'_H=3000\mev$ in agreement with previous works. For details on this procedure see Refs.~\cite{Gamermann:2006nm,Gamermann:2007fi}.

%----------------------------------
% Relevant Channels and amplitudes
%----------------------------------

%----------------------------------
% J/psi pi 
%----------------------------------

In the hidden charm sector with $I=1$ and $J=1$ the three channels contribute: $\jpsi\pi$, $\eta_c\rho$ and $D\bar D^*+c.c.$. 
In all sectors we will neglect channels with hidden strangeness, such as $e.g.$ $D_s\bar K$, as their thresholds are typically around $500\mev$ higher than the energies we are interested in.
The corresponding coupled channel scattering potential can be written as
\begin{equation}\label{Eq:VJpsi}
 V_{ij}(s,t,u) = - \frac{\xi _{ij}}{4f^2}(s-u) \epsilon\cdot \epsilon', \qquad
 \xi _{ij} = \left( \begin{array}{ccc}  0 & 0 & \sqrt{8/3}\gamma \\  0 & 0 & \sqrt{8/3}\gamma \\  \sqrt{8/3}\gamma & \sqrt{8/3}\gamma & -\psi   \end{array}\right)
\end{equation}
Note the absence of any interactions between the channels $\jpsi\pi$ and $\eta_c\rho$. All non-vanishing interactions involve a charmed meson pair at least. 

%----------------------------------
% Dstar pi 
%----------------------------------
In the vector-pseudoscalar scattering with open charm we find three channels, $D^*\pi$ and $D^*\eta$ for isospin $I=1/2$ and just $D^*\pi$ for $I=3/2$.
The resulting potentials can be parametrized in the same fashion:
\begin{equation}\label{Eq:V2}
 V_{ij}(s,t,u) = - \frac{\xi _{ij}}{4f^2}(s-u) \epsilon\cdot \epsilon', \qquad
 \xi _{ij} = \left( \begin{array}{ccc}  -2 & 0 & 0 \\  0 & 0 & 0 \\  0 & 0 & 1   \end{array}\right) \ ,
\end{equation}
where the 2x2 upper-left box corresponds to the isospin $I=1/2$ case and the 1x1 lower-right one is for $I=3/2$. Note, however, that
in the absence of any $D^*\eta$ interactions the potentials decouple and we effectively find two single-channel potentials, one for each isospin case.

%----------------------------------
% D pi 
%----------------------------------
The situation is slightly different in the case of charmed pseudoscalar mesons. 
For open charm $C=1$ and total angular momentum $J=0$ we find the channels $D\pi$ and $D\eta$ for Isospin $I=1/2$ and just $D\pi$ for $I=3/2$.
The amplitude for pseudoscalar pseudoscalar scattering has a slightly different form, as the off-diagonal interactions are non-vanishing. 
 The full amplitude reads 
\begin{equation}
% \frac{1}{12f^2, 
V^{I=1/2}_{D\pi,D\eta}(s,t,u)\left(
\begin{array}{cc}
 \frac{-2 (2 \gamma +1) m_D^2-2 (2 \gamma +1) m_{\pi }^2-4 s+4 u+s \gamma +5 u \gamma }{12 f^2} & -\frac{(4 \gamma +2) m_D^2+2 \gamma  m_{\eta }^2+2 \gamma  m_{\pi }^2+2 m_{\pi }^2-3 s \gamma -3 u
   \gamma }{12 f^2}  \\
 -\frac{(4 \gamma +2) m_D^2+2 \gamma  m_{\eta }^2+2 \gamma  m_{\pi }^2+2 m_{\pi }^2-3 s \gamma -3 u \gamma }{12 f^2} & \frac{-2 (2 \gamma +1) m_D^2-4 \gamma  m_{\eta }^2-2 m_{\pi }^2+3 s \gamma +3 u
   \gamma }{36 f^2}    
\end{array}
\right).
\end{equation}
for $D\pi$ and $D\eta$ in  $I=1/2$ and
\begin{equation}
 V_{D\pi}^{I=3/2}=\frac{-(2 \gamma +1) m_D^2-(2 \gamma +1) m_{\pi }^2+s-u+2 s \gamma +u \gamma }{6 f^2}
\end{equation}
for $D\pi$ in $I=3/2$. 
However, the fact that $\gamma\ll1$ allows us to simplify this expression to some extent, which can be written in the same compact form as for the vector-pseudoscalar case:
\begin{equation}
 V(s,t,u) = \frac{s-u}{6 f^2}\left(
\begin{array}{ccc}
 -2 & 0 &0 \\
 0 & 0 &0\\
   0&0& 1
\end{array}
\right) 
- \frac{m_D^2+m_{\pi }^2}{18 f^2}\left(
\begin{array}{ccc}
 3 & 3 & 0 \\
 3 & 1 & 0\\
 0 & 0 & 3
\end{array}
\right) +\mathcal{O}(\gamma) \ ,
\end{equation}
where the terms proportional to $\gamma$ are now cobtained in $\mathcal{O}(\gamma)$. This is similar in form to the case for charmed vector mesons, up to the second part induced by the mass term in the purely pseudoscalar Lagrangian. 
%----------------------
% Unitarized Amplitude
%----------------------

The corresponding $S$-wave projections will be used as the kernel for the Lippmann-Schwinger equation depicted in Fig.~\ref{Fig:diagram}(a). 
Using the on-shell formalism this simplifies to a simple algebraic equation that can easily be solved as
\begin{equation}\label{Eq:T}
 T= (1 - VG)^{-1}V \vec \epsilon\cdot \vec \epsilon' \ ,
\end{equation}
for the scattering of vector mesons off pseudoscalar ones, while for the purely pseudoscalar case one
simply needs to remove the polarization vectors from the previous equation.
The diagonal matrix $G$ contains the two-meson loops which can be calculated analytically as
\begin{eqnarray}\label{Eq:G_Vacuum}
 G_{ii}(s) &=& {\rm i} \int\!\frac{\mathrm d^4q}{(2\pi)^4} \frac{1}{[q^2-m_1^2+{\rm i}\varepsilon][(P-q)^2-m_2^2+{\rm i}\varepsilon]}\\
 &= &{\frac{1}{16\pi ^2}}\biggr( \alpha_i+\log{\frac {m_1^2 }{ \mu ^2}}+{\frac{m_2^2-m_1^2+s}{ 2s}}\log{\frac{m_2^2}{ m_1^2}}+\nonumber\\ 
  &&{\frac{p}{\sqrt{s}}}\Big( \log{\frac{s-m_2^2+m_1^2+2p\sqrt{s} }{ -s+m_2^2-m_1^2+2p\sqrt{s}}}+\log{\frac{s+m_2^2-m_1^2+2p\sqrt{s} }{ -s-m_2^2+m_1^2+  2p\sqrt{s}}}\Big)\biggr)
\end{eqnarray}
where the index $i$ refers to the pair of meson with masses $m_1$ and $m_2$ and the center of mass momentum is given by $P\,^2 = s$. 
The loop integrals are calculated using dimensional regularization. 
We will fix the free parameters following Ref.~\cite{Gamermann:2007fi}, $e.g.$ the subtraction constant is $\alpha\,_H=-1.55$ 
at the scale $\mu=1.5\gev$ to reproduce the $D^*_{s0}(2317)$ in the open charm and open strangeness sector with quantum numbers  $J^{P}=1^{+}$ and $I=0$. 

%===============================================================
\subsection{Interactions at finite temperatures}\label{Sec:itf}
%===============================================================
%------------------------------------------------------------------------------------------------------------------------------------------------------------------------
% Figure Feynman Diagrams Self Consistency 
%------------------------------------------------------------------------------------------------------------------------------------------------------------------------
\begin{figure}[t]
\centering
\includegraphics[width=.7\linewidth]{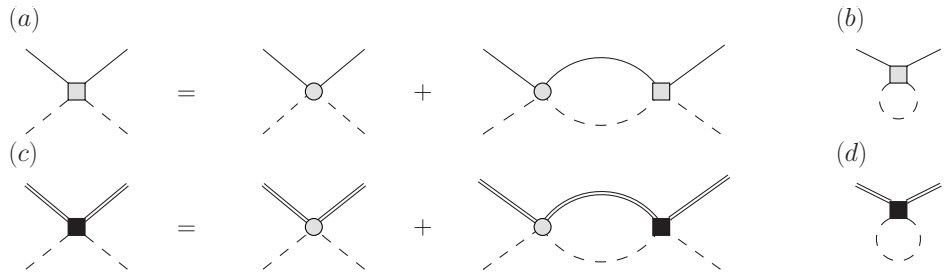}
\caption{Feynman Diagrams representing the Lippmann-Schwinger equation for the meson-pion interaction at finite temperature (a), the resulting self-energy from closing the pion line (b), the in-medium meson-meson interaction obtained with dressed propagators (c), and the resulting meson self-energy which needs to be obtained selfconsistently (d). } \label{Fig:diagram}
\end{figure}
%------------------------------------------------------------------------------------------------------------------------------------------------------------------------
In this section we will outlay the technical framework we are using to incorporate the influence of a hot pion bath on the open and hidden charm mesons.
First, we will use the Imaginary Time Formalism (ITF) to study the impact of finite temperatures. The technical details will be discussed later. 
Essentially, ITF is a framework to modify the propagators while the interaction vertices remain unchanged. 
Secondly, we will dress the propagators for the mesons $M$ with the self energy obtained from closing the pion line in the $M\pi\to M\pi$ scattering amplitude, {\it cf.} Fig.~\ref{Fig:diagram} (b).
This way the dressed propagators (double lines) account for the pion bath. 
Since the unitarized amplitude calculated using dressed propagators (Fig.~\ref{Fig:diagram}(c)) in turn modifies the self energy (Fig.~\ref{Fig:diagram}(d)) the procedure needs to be iterated several times to ensure self consistent results.

We will now review the essential details of ITF, a more comprehensive study can be found {\it e.g.} in Refs.~\cite{galekapustabook,lebellac}. 
In short, the zeroth components of the four-momenta are modified using
\begin{equation}
 q_0\to {\rm i}\omega_n = {\rm i} 2\pi nT, \qquad \int\!\frac{\mathrm d^4q}{(2\pi)^4}\to {\rm i}T\sum_n \int\!\frac{\mathrm d^3q}{(2\pi)^3}
\end{equation}
with the discrete Matsubara frequencies $\omega_n = {\rm i}2\pi nT$. 
Once the sums over the Matsubara frequencies are performed one may analytically continue the result to external frequencies $\omega+{\rm i}\varepsilon$. 
In the following we will lay out how this affects the loops and self energies. 

%-------------
%    Loops  
%-------------
We will start by calculating the meson-meson loop at finite temperature. Applying  the ITF Feynman rules we find 
\begin{equation}\label{Eq:loop1}
 G_{MM'}(W_m,\vec p;T) = - T\sum_n \intq D_M(\omega_n,\vec q;T)D_{M'}(W_m-\omega_n,\vec p-\vec q;T).
\end{equation}
where the propagator for a meson $M$ in a hot medium is given by
\begin{equation}
 D_M(\omega_n,\vec q;T) = [({\rm i}\omega_n)^2-\vec q\,^2-m_M^2-\Pi_M(\omega_n,\vec q;T)]^{-1}
\end{equation}
To carry out the Matsubara sum it is convenient to use the spectral (Lehmann) representation
\begin{equation}
 D_M(\omega_n,\vec q;T) = \int\!\mathrm d\omega \frac{S_M(\omega,\vec q;T)}{{\rm i}\omega_n-\omega} \ ,
\end{equation}
which linearizes the $\omega_n$ dependence in the denominator, where the spectral function is given by 
\begin{equation}
 S_M(\omega_n,\vec q;T) = -(1/\pi)\mathrm{Im}(D_M(\omega_n,\vec q;T))\,.
\end{equation}
Simplified this way we can carry out the sum over the discrete frequencies in Eq.~(\ref{Eq:loop1})  as
\begin{equation}
T \sum_n \frac{1}{[{\rm i}\omega_n-\omega][{\rm i}W_m-{\rm i}\omega_n-\Omega]} = - \frac{1+f(\omega,T)+f(\Omega,T)}{{\rm i} W_m- \omega-\Omega}
\end{equation}
with $f(\omega,T)=[\exp(\omega/T)-1]^{-1}$ the meson Bose distribution function at temperature $T$. 
Extrapolating to the real axis with the replacement ${\rm i}W_m\to p^0+{\rm i}\varepsilon$, Eq.~(\ref{Eq:loop1}) then becomes:
\begin{eqnarray}\label{Eq:G_TVac}
 G_{MM'}(p^0,\vec p;T) = \intq \int\!\mathrm d\omega\int\!\mathrm d\Omega \frac{S_M(\omega,\vec q;T)S_{M'}(\Omega,\vec p - \vec q;T)}{p^0-\omega-\Omega+{\rm i}\varepsilon}     [1+f(\omega,T)+f(\Omega,T)] \ .
\end{eqnarray}
This result holds true for an arbitrary two-meson loop at finite temperatures. 
However, this could be further simplified in the absences of self energy. 
For example, the interactions of a pion with a dominantly pionic medium are weak and the resulting self energies vanish. 
In this case the spectral function becomes the delta distribution:
\begin{equation}
 S_M(\omega,\vec q;T)\to \frac{\omega_M}{\omega}\delta(\omega^2-\omega_M^2)
\end{equation}
with $\omega_M=\sqrt{q_M^2+m_M^2}$, allowing us to carry out one of the spectral integrals.  This prescription will be used for all pion spectral functions throughout this work.

At this point some comments on the regularization of the loop in the hot pionic medium are in order. 
We will use a sharp cutoff to regularize the three-momentum integration here. 
However, to be consistent with Eq.~(\ref{Eq:G_Vacuum}) we will follow the approach by Tolos et al. in Ref.~\cite{Tolos:2009nn}. 
We will take the original calculation, performed in dimensional regularization and without temperature dependence, $G(s)$ from Eq.~(\ref{Eq:G_Vacuum}), and correct it by a temperature-dependent and pion-medium modified function $\delta G(s,T)$ defined as
\begin{equation}\label{Eq:G_Correction}
 \delta G (s,T) = \lim_{\Lambda\to\infty} \delta G_\Lambda (s,T) = \lim_{\Lambda\to\infty} [ G_\Lambda (s,T) - G_\Lambda (s,T=0) ]
\end{equation}
Using $G(s,T)=G(s)+\delta G (s,T)$ allows us to smoothly continue from $T=0$ to the finite temperature case. 
At the same time, this prescription drastically reduces the cutoff dependence of the results. 

%------------------
%   Self Energy
%------------------

The second important temperature-dependent quantity is the self energy of a meson $M$ obtained from closing the pion line in the $M\pi\to M\pi$ $T$-matrix, see Figs.~\ref{Fig:diagram}(b) and (d).
Applying  the ITF Feynman rules we find
\begin{equation}
 \Pi_{M}(W_m,\vec p) = T\int\!\frac{\mathrm d^3q}{(2\pi)^3}\sum_n D_\pi(\omega_n,\vec q;T)T_{M\pi}(\omega_n+W_m,\vec p+\vec q;T) \ .
\end{equation}
We deal with the Matsubara frequencies in the same way as for the loop, {\it i.e.} by introducing spectral representations for both the pion propagator and the $M\pi$ $T$-matrix.
This allows us to carry out the Matsubara sum and, using the corresponding delta distribution for the bare pion, we find for real energies $p^0$
\begin{eqnarray}\label{Eq:Pi}
 \Pi_M(p^0,\vec p) = \int\!\frac{\mathrm d^3q}{(2\pi)^3} \int\!\mathrm d\Omega
 \frac{ f(\Omega,T)-f(\omega_\pi,T)}{(p^0)^2 - (\omega_\pi-\Omega)^2 + {\rm i}\varepsilon}  
 \left(-\frac1\pi\right) \mathrm{Im} T_{M \pi}(\Omega,\vec p+\vec q;T). 
\end{eqnarray}
%===========================================================
\section{Results}\label{Sec:results}
%===========================================================
In the following we will present the results of applying the theoretical framework laid out in the previous section.
We will begin by studying the effect unitarization has on the cross section for elastic and inelastic $\jpsi\pi$ scattering. 
Then we follow with the study of the width and spectral functions of open and hidden charm mesons induced by a hot pion bath.
%===========================================================
\subsection{Cross Sections}\label{Sec:crosssection}
%===========================================================
%------------------------------------------------------------------------------------------------------------------------------------------------------------------------
% Figure Cross Sections J/psi pi -> Jpsi pi, eta_c rho and DD*
%------------------------------------------------------------------------------------------------------------------------------------------------------------------------
%------------------------------------------------------------------------------------------------------------------------------------------------------------------------
%---------------------------------------------
% Discussion Cross Section Jpsi pi -> Anything
%---------------------------------------------
In this section we calculate the cross sections for $\jpsi\pi$ scattering into different final states, elastic and inelastic, using the theoretical framework introduced in Sec.~\ref{Sec:su4}.
This is instructive for several reasons. First of all, it gives us an idea about the order of magnitude for the interactions of the $\jpsi$ in a pionic medium. 
The inelastic cross section in particular allows us to gauge the hadronic dissociation of the $\jpsi$ in such a scenario.

\begin{figure}[ht]
\centering
\begin{minipage}{0.475\linewidth}
 \includegraphics[width=\linewidth]{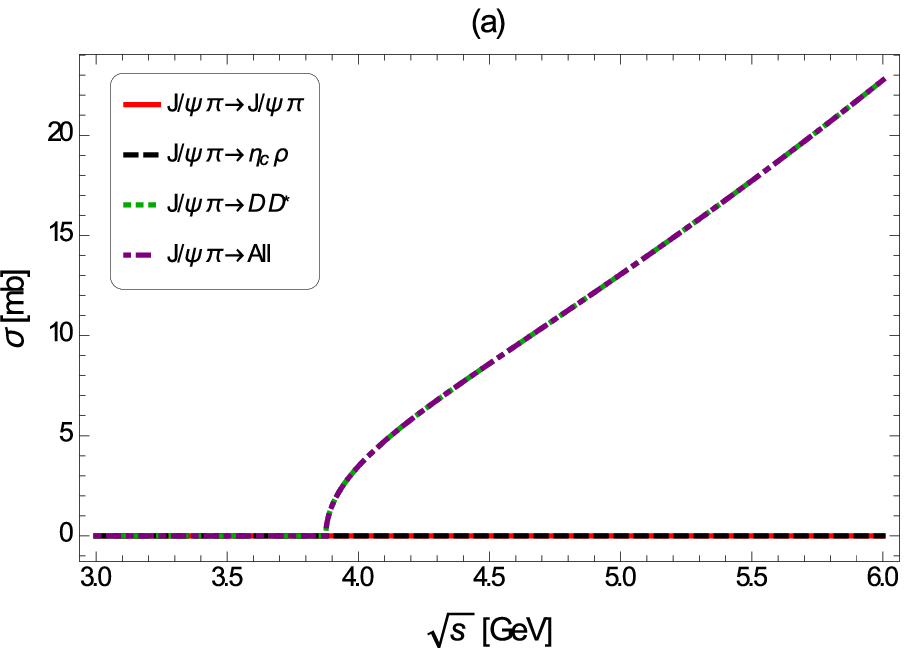}
\end{minipage}
\begin{minipage}{0.475\linewidth}
 \includegraphics[width=\linewidth]{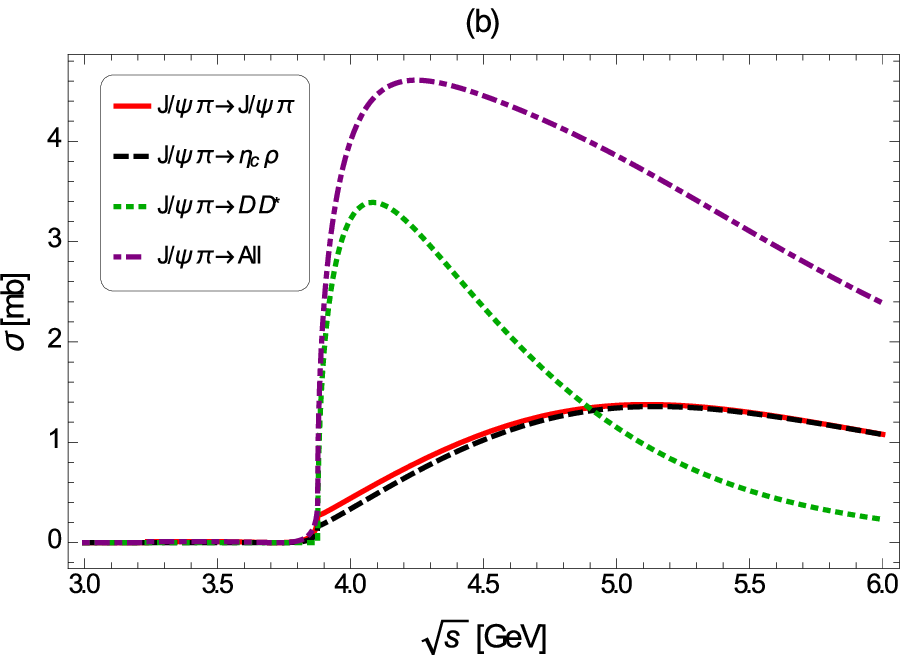}
\end{minipage}
\caption{Cross Sections for $J/\psi\pi\to X$ employing the non-unitarized amplitudes (a) and the unitarized ones (b).} \label{Fig:JpsipiX_Vacuum}
\end{figure}

Fig.~\ref{Fig:JpsipiX_Vacuum} shows the cross sections for the processes $J/\psi\pi \to J/\psi\pi$, $J/\psi\pi \to \eta_c\rho$ and $J/\psi\pi \to DD^*$. 
The left panel displays the cross sections using non-unitarized amplitudes, the right panel shows the cross section using the unitarized amplitudes according to Eq.~(\ref{Eq:T}).
When comparing these two we see several noteworthy differences. 

First of all, the vanishing amplitude for $J/\psi\pi \to J/\psi\pi$ and $J/\psi\pi \to \eta_c\rho$ at tree level means that the corresponding cross sections vanish as well. 
The lack of $J/\psi\pi \to J/\psi\pi$ scattering can be understood as an immediate consequence of the interpretation of the pion as the Goldstone boson of QCD.
However, once the amplitude is unitarized, these processes can occur through charmed meson loops which give rise to a  cross section in the range of $1~\mathrm{mb}$. 
This will also become an important feature once we consider in-medium effects as we will see in the following section. 

The second interesting observation when studying the impact of unitarization is that in the case of $\jpsi\pi$ going into a charmed meson pair the high-energy behaviour changes significantly. 
Whereas the non-unitarized amplitude produces a cross section that linearly rises from $4\gev$ onwards, the cross section in the unitarised case peaks shortly after this point and begins to drop again approaching zero around $6\gev$. 

%---------------------------------------------
% Comparison to other works
%---------------------------------------------
Although slightly different in shape, the size of our unitarized cross sections is in line with what has been found with the chiral phenomenological models developed in  \cite{Haglin:2000ar,Bourque:2008ta}, the extended Nambu-Jona-Lasinio model \cite{Bourque:2008es},  a meson-exchange model \cite{Oh:2000qr}, a non-relativistic quark potential model \cite{Wong:2001td}, or a QCD-sum-rule approach \cite{Duraes:2002px}.
The difference in shape stems from the effect a unitarized amplitude has on the cross section at high energies as was discussed before. 
This highlights the importance of using this more modern approach and the need of updating the calculation on other quantities such as the spectral function accordingly, as we will do in the next sections.

%======================================================
\subsection{Temperature Effect on Loops and T-Matrices}
%======================================================
% \subsubsection{Loops}
%======================================================

We will start with the discussion of the hot pion bath effects by studying the  two-body loops for the channels most relevant to our discussion: 
$D\pi$, $D^*\pi$, $\jpsi\pi$ and $DD^*$. 
The effect of non-vanishing temperatures is best studied in the imaginary parts as here the left and right-handed cuts are most obvious.

\begin{figure*}[th]
\centering
  \includegraphics[width=\linewidth]{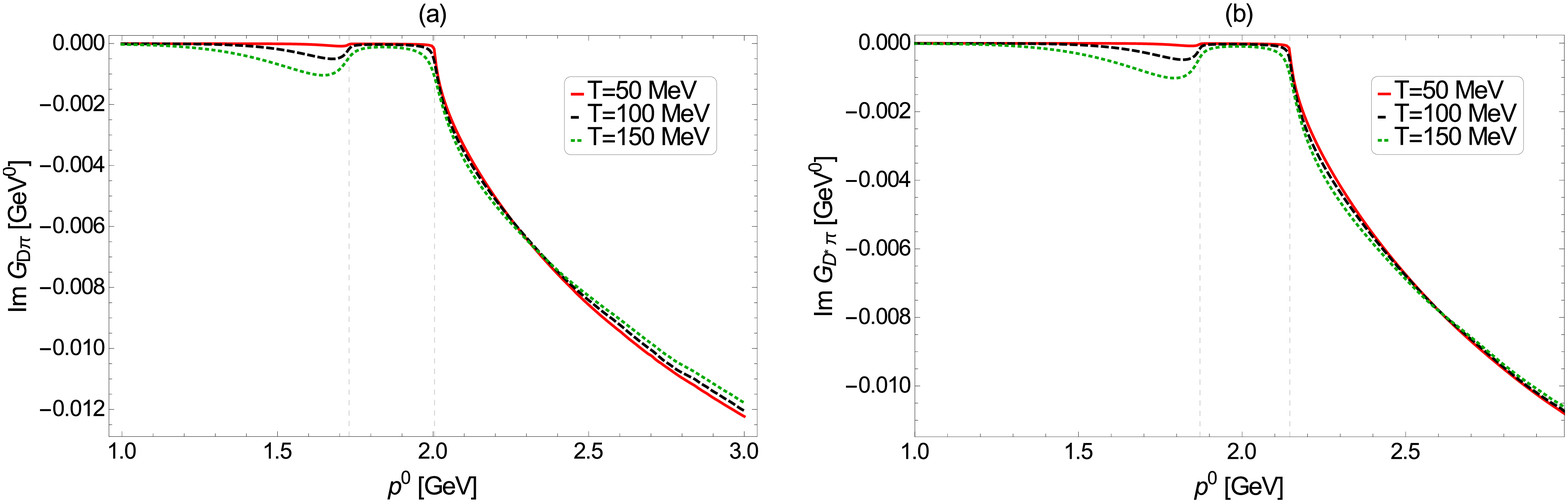}\\ %\vspace{-3cm}
  \includegraphics[width=\linewidth]{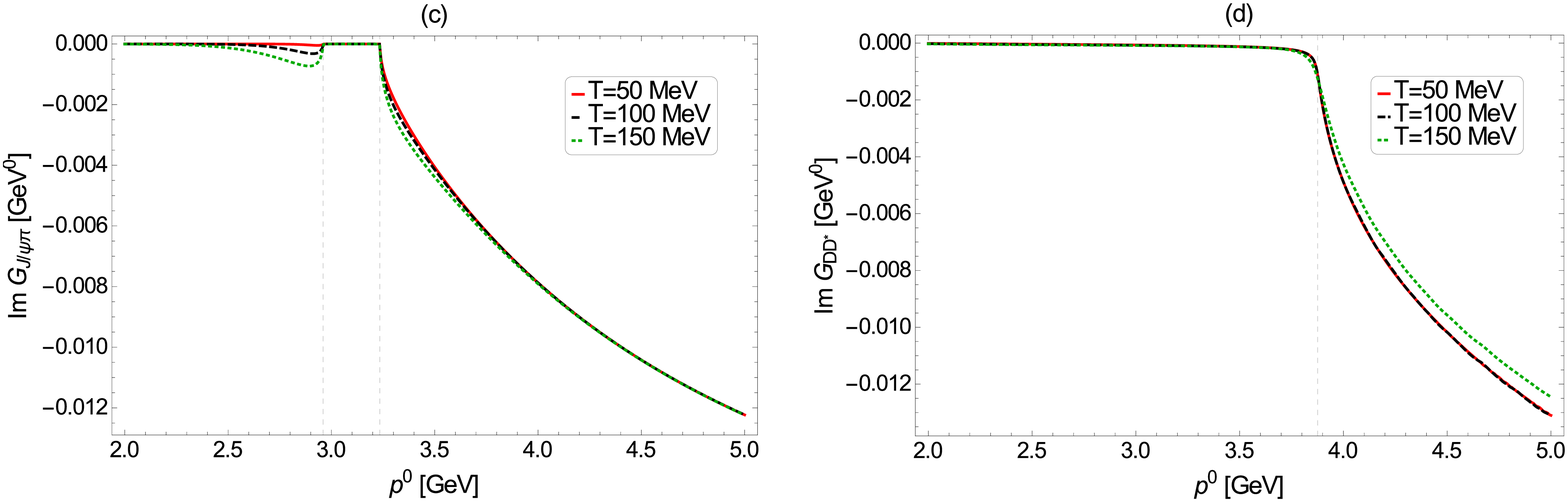}
      \setlength{\unitlength}{0.01\linewidth}
    \begin{picture}(100,0)    % picture environment for inset
        \put(59,10){\includegraphics[width=22\unitlength]{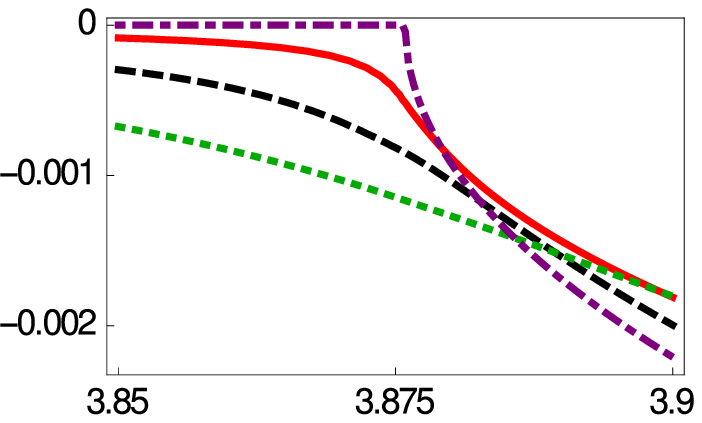}}
    \end{picture}

\caption{
Imaginary part of the meson-meson loop as a function of the energy $p^0$ at temperatures $T = 50$, 100, 150 MeV and $\vec{p} = 0$. (a): $D\pi$, (b): $D^*\pi$, (c): $\jpsi\pi$ and (d): $D D^*$. The vertical dashed lines correspond to $\mid m_1-m_2\mid$ and $m_1 + m_2$. The $T = 0$ case (dot-dashed line) is also shown for reference in the inset of the $D D^*$ loop panel.
} \label{Fig:Loop}
\end{figure*}

In Fig.~\ref{Fig:Loop} we see the imaginary parts of the relevant loops. We see a left-hand cut from $|m_1-m_2|$ and a right-hand cut from $m_1+m_2$. 
The left-hand cut, sometimes referred to as Landau-cut, exists only for finite temperatures, the right-hand cut is the usual unitarity cut plus a correction. 
% -------------------------------------------------------------
% -------------------------------------------------------------

If we take a closer look at the impact of temperature and the dressing of the mesons on the loops, we notice certain similarities and differences between the different channels. 
Most importantly, in all cases that involve a pion -- $D\pi$, $D^*\pi$ and $\jpsi\pi$ -- the loop is very similar and basically differs only by a shift according to the other meson's mass. 
This is due to the fact that the Bose-Einstein factors at given temperature produce the largest contributions for the smallest mass. 
So the effect of the temperature is dominated by the contribution from the pion bath. 
At the same time, due to the small mass of the pion, the gap between $m_1-m_\pi$ and $m_1+m_\pi$ with $m_1$ the mass of the non-pion in the loop is also small and a large part of the energy range obtains sizable corrections. 
This is best highlighted when contrasted with the case of the two charmed mesons, $D$ and $D^*$. 
Here, the two masses are both of the order of $2\gev$ and hence one order of magnitude larger than the highest temperatures studied here.
This leads to a suppression of the corresponding Bose-Einstein factors that only becomes a significant contribution when one increases the temperatures to the order of the mass and thus well out of the regime that is sensible in the context of this work. 
Moreover, the proximity of the masses to each other means that the left-hand cut is moved to energies close to zero. Note, however, that the correction of the $DD^*$ loop due to temperature is quite noticeable close to threshold, as can be better appreciated in the figure inset, where the $T=0$ case is also shown for reference. This is due to the substantial modification of the $D\pi$ and $D^*\pi$ interactions in a hot medium where pions constitute essentially the major component.

%==================================================
% \subsubsection{T-Matrices}
%==================================================
% -------------------------------------------------------------
\begin{figure*}
\centering
  \includegraphics[width=\linewidth]{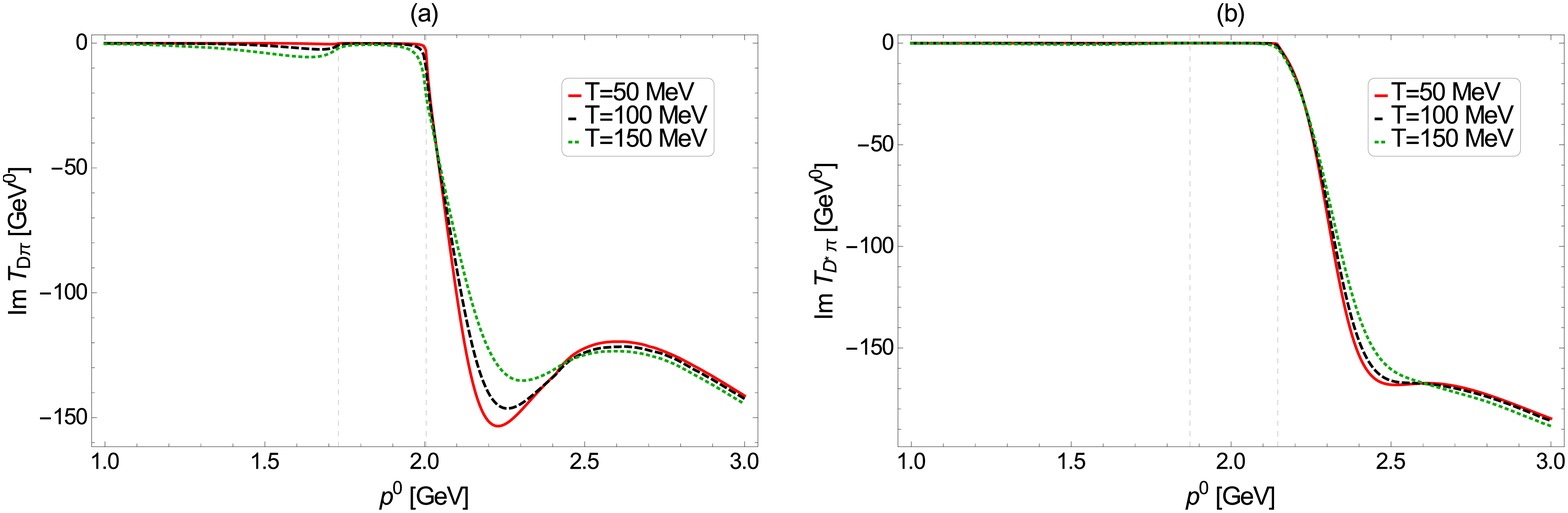}\\ %\vspace{-3cm}
  \includegraphics[width=\linewidth]{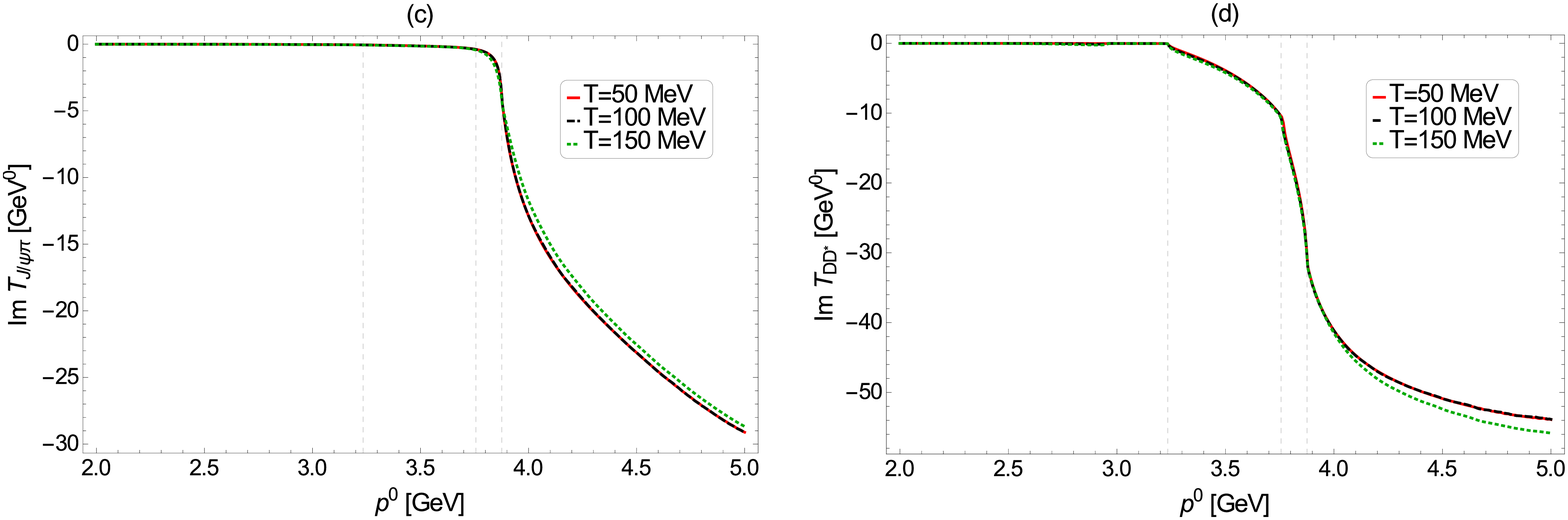}
\caption{
Imaginary part of the unitarized amplitude as a function of the energy $p^0$ at temperatures $T = 50$, 100, 150 MeV and $\vec{p} = 0$. (a): $D\pi \to D\pi$, (b): $D^*\pi \to D^*\pi$, (c): $\jpsi\pi \to \jpsi\pi$ and (d): $D D^* \to D D^*$.  The vertical dashed lines correspond to $\mid m_1-m_2\mid$ and $m_1 + m_2$ for for panels (a) and (b), and  $m_{J/\psi}+m_\pi$, $m_{\eta_c}+m_\rho$ and $m_D + m_{D^*}$ for panels (c) and (d).
} \label{Fig:T}
\end{figure*}
% -------------------------------------------------------------
In Fig.~\ref{Fig:T} we see the imaginary parts of the relevant T-matrices. Now, as was mentioned before, in ITF only the propagators are modified, the interaction potentials remaining unchanged. 
So any change in the unitarized amplitude should in principle reflect the change in the two-body loop as was discussed before. 
However, there are several points worth noting here. 
%-------------
% D(*)pi
%-------------
When we look at the charmed mesons unitarized $D\pi$ and $D^*\pi$ amplitudes (Fig.~\ref{Fig:T}, upper subplots), two things attract attention.
First, there is a bump in the imaginary part for $D\pi$ that is not so pronounced for its vector partner and, secondly, there is a larger contribution from the left-hand cut in the case of $D\pi$.
Given that the loops are identical, up to the shift due to the different masses, the reason for the differences in the amplitudes comes obviously from the scattering potential. 
This difference stems from the Lagrangians used in Eq.~(\ref{Eq:lag}), with a mass term contributing only to the scattering of two pseudoscalar mesons. This leads to a constant shift of the potential and also induces off-diagonal elements, mixing the $D\pi$ and $D\eta$ channels in $I=1/2$, that are absent for the charmed vector meson. 
As a consequence, the off-diagonal interactions that without the mass term are suppressed, become more significant and enhance the effect of the $D\eta$ channel.
We also note a difference between the $D\pi$ and $D^*\pi$ scattering amplitudes below the threshold. 
While there is a small bump for the $D\pi$ case we see none in the case of $D^*\pi$. 
The reason is that in this region the potential $V$ for the $D\pi$ case is significantly larger than for the $D^*\pi$ case because of the shift induced by the mass term.
Therefore, while the potentials are almost identical for the two cases elsewise, the mass term of the purely pesudoscalar  Lagrangian induces visible changes in the $T$-matrix and so it breaks heavy quark spin symmetry to some extent.  
%-------------
% Jpsi pi
%-------------
When we compare the $D\pi, D^*\pi$ unitarized amplitudes with those corresponding to the two most important channels with hidden charm -- $D\bar D^*+c.c.$ and $\jpsi\pi$ -- the first thing one notices is that, in the hidden charm case, they are  smaller by a factor $4-5$. 
This is easily understood when one looks at the potential given in Eq.~(\ref{Eq:VJpsi}): all contributions in this case are suppressed by factors $\gamma$ or $\psi$. 

What is particularly interesting about the unitarized amplitudes here is how they are affected by the opening of the different channels. 
If we neglect the $\eta_c\rho$ contributions for a moment -- which is sensible as their impact is indeed rather small -- we find the following. 
With the vanishing potential for direct $\jpsi\pi$ scattering the coupled-channel interaction ({\it cf.} Eq.~(\ref{Eq:VJpsi}) and the discussion of the cross section) goes exclusively through the $DD^*$ loop and hence this threshold and not the $\jpsi\pi$ threshold itself dominates the imaginary part of the T-matrix. 
In the case of $DD^*\to DD^*$ which can also go through $\jpsi\pi$ loops we see the opposite, an imaginary part opening up below the nominal threshold.

Given that the unitarized amplitude for $\jpsi\pi$ scattering is dominated by the coupling to the $DD^*$ channel a natural thing to study is the impact of using the spectral functions for the charmed mesons shown in Fig.~\ref{Fig:SpecFunc}. Consistenly which the effects observed in the $DD^*$ loop, the changes in the   $\jpsi\pi$ amplitude  are mainly concentrated at threshold, and their dependence on temperature is quite mild.

%=============================================
\subsection{Self Energy and Spectral Function}
%=============================================
% \subsubsection{Self Energy}
%=============================================
%---------------------------------------------------------
\begin{figure*}
\centering
 \includegraphics[width=\linewidth]{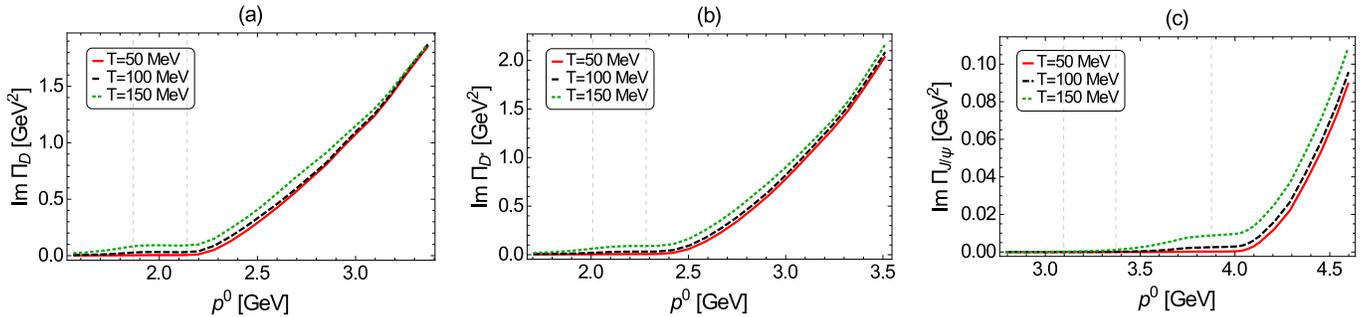}
\caption{
Imaginary part of the pion-induced self energy as a function of the energy $p^0$ at temperatures $T = 50$, 100, 150 MeV and $\vec{p} = 0$. 
(a): $D$, (b): $D^*$ and (c): $\jpsi$. The vertical lines in panels (a) and (b) indicate $p^0 = m_{D^{(*)}}$ and $p^0 = m_{D^{(*)}} + 2 m_\pi$, whereas the ones in panel (c) mark $m_{\jpsi}$, $m_{\jpsi}+2m_\pi$ and $m_D + m_{D^*} + m_\pi$.
}
 \label{Fig:SelfEnergy}
\end{figure*}
%---------------------------------------------------------
In the following we will discuss the imaginary part of the self energies for the mesons in question. The real part enters the spectral function as 
\begin{equation}
 Re\big(\bar \Sigma(p^0,\vec p;T)\big) = Re\big( \Sigma(p^0,\vec p;T) - \Sigma(p^0,\vec p;T=0)\big)
\end{equation}
in order to obtain the correct limit for temperatures approaching zero, $i.e.$ the mass becomes the vacuum mass. 
However, it turns out that in the cases here this difference accounts for little more than numerical noise compared to the vacuum mass. 
Hence, in Fig.~\ref{Fig:SelfEnergy} we only show the imaginary parts of the self energies of $D$, $D^*$ and $\jpsi$ resulting from their interaction with the hot pion bath. 
We notice that the charmed mesons differ only little in comparison to each other. 
In both cases, we observe a temperature induced increase around the mass of the respective meson followed by a sharp rise starting at $m_{D^{(*)}}+2m_\pi$. 
The latter corresponds to the energy of the meson at rest from which this decay becomes possible.

In the case of the $\jpsi$, the self energy is first of all significantly smaller, being more than an order of magnitude smaller in comparison to the charmed meson values. Secondly, as the $\jpsi\pi$ diagonal amplitude is much suppressed, the sharp rise is not seen at $m_\jpsi + 2 m_\pi$ but rather at  $m_D + m_{D^*}+m_\pi$, an excitation that is accessed through the non-diagonal $\jpsi\pi \to D D^*$ amplitude. 

%=============================================
% \subsubsection{Spectral Function}
%=============================================

%------------------------------------------------------
\begin{figure*}
\centering
 \includegraphics[width=\linewidth]{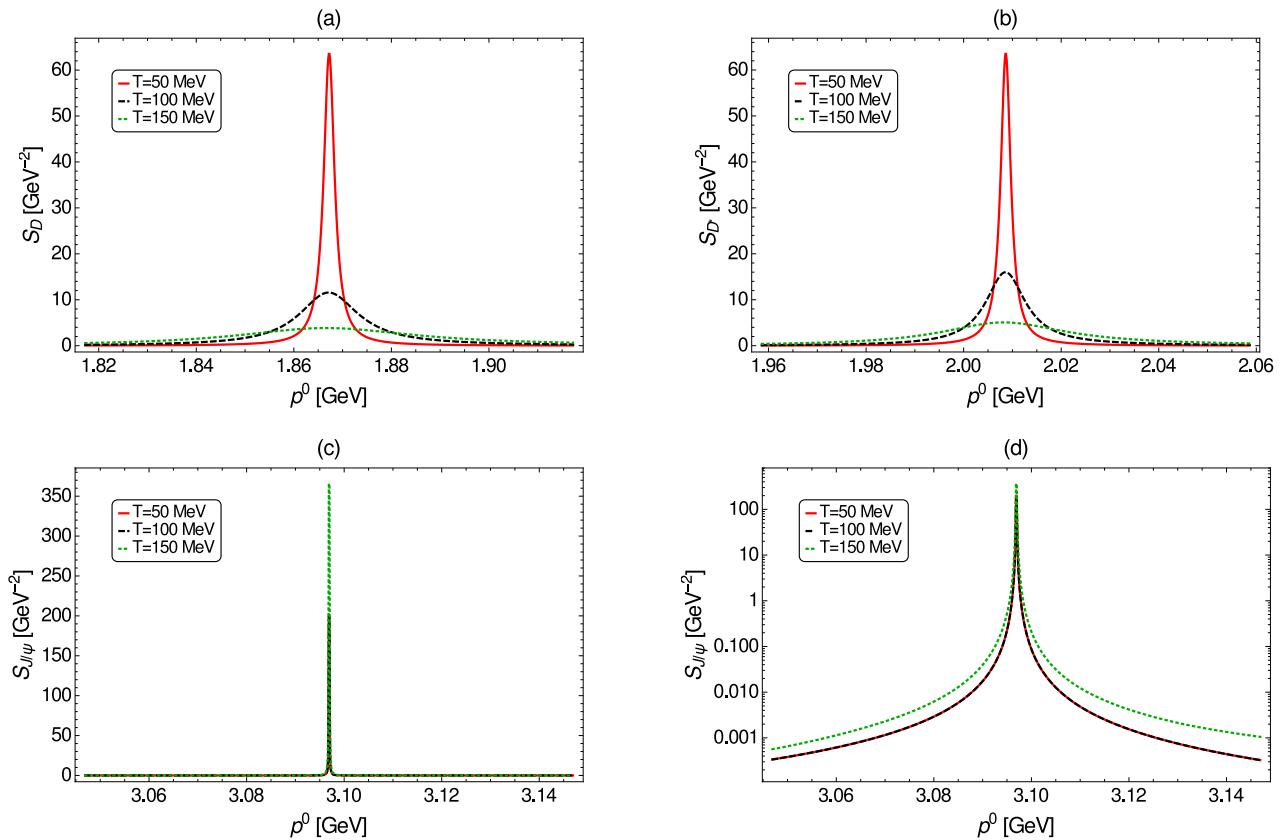}
\caption{
Meson spectral function as a function of the energy $p^0$ at temperatures $T = 50$, 100, 150 MeV and $\vec{p} = 0$. a): $D$, (b): $D^*$, (c): $\jpsi$ and (d): $\jpsi$ in a logarithmic scale.
} \label{Fig:SpecFunc}
\end{figure*}
%------------------------------------------------------

In Fig.~\ref{Fig:SpecFunc} we show the resulting spectral functions as a function of the energy at increasing temperatures $T=50,100,150\mev$.
Naturally, the behaviour of the spectral functions mirrors the behaviour of the self energy. 
In the case of the charmed mesons we see how the strength spreads considerably when increasing the temperature from 50 to 150~MeV. 
In other words, the spectral function of the charmed mesons broadens significantly in a finite-temperature pion bath. 

The situation is obviously different for the $\jpsi$ spectral function.
Notice that in this case we have also included the $\jpsi$ vacuum width of $\Gamma_{\jpsi}^\mathrm{Vac}=(93\pm3)\kev$~\cite{Olive:2016xmw}.
On one hand this makes the numerical determination of the spectral function easier, on the other hand it gives an intuitive measure of the impact of the pion-induced width.
The large self-energy for the charmed mesons made this incorporation irrelevant there. 
The identical $\jpsi$ spectral functions seen for the temperatures $T=50\mev$ and $T=100\mev$ mean that the pion-induced width is suppressed compared to the vacuum width. One needs to go to temperatures as large as
$T=150\mev$ to start seing noticeable broadening of the $\jpsi$ spectral function compared to its free case shape, as is better seen in the logarithmic plot of Fig.~\ref{Fig:SpecFunc}.

The reasons for this are basically two-fold. 
First and foremost, as was discussed in the previous sections, the interactions that contribute to the $\jpsi$ are weak. 
This results in a unitarized amplitudes (Fig.~\ref{Fig:T}) that are suppressed by at least one order of magnitude compared to the charmed mesons. 
Moreover, as was discussed in the previous section, the vanishing leading interaction for  $\jpsi\pi\to\jpsi\pi$ leads to the imaginary part of 
the $T$-matrix effectively opening at the $DD^*$ threshold (compare Fig.~\ref{Fig:T}).
This is a significantly higher energy and thus further away from the mass of the $\jpsi$. 
Both factors combined lead to the result we are seeing, namely that the modification of the $\jpsi$ spectral function is suppressed compared to the ones for the charmed mesons.

In a previous work, Gale and Haglin calculated the spectral function of the $\jpsi$  \cite{Haglin:2000ar}. 
While similar in shape, their results are about two orders of magnitude larger than our findings. 
Part of this is trivial due to a difference of $2\pi$ in the definition of the spectral function. 
However, this means that our spectral function is significantly narrower due to the use of improved techniques developed in the meantime. 
While a broadening spectral function for $T=150\mev$ can be seen from our results, it still remains similar to the vacuum case.
From this we can conclude that the $\jpsi$ retains most of its shape in hot pionic matter.
A suppession in the detection of $\jpsi$ mesons in a hot pionic medium should therefore not be attributed to the interaction with the surrounding hadrons. 
Thus, if observed, such a drop might indeed point towards the detection of a QGP.

%==================================================
% \subsubsection{Width as a function of Temperature}
%==================================================

\begin{figure*}
\centering  
  \includegraphics[width=\linewidth]{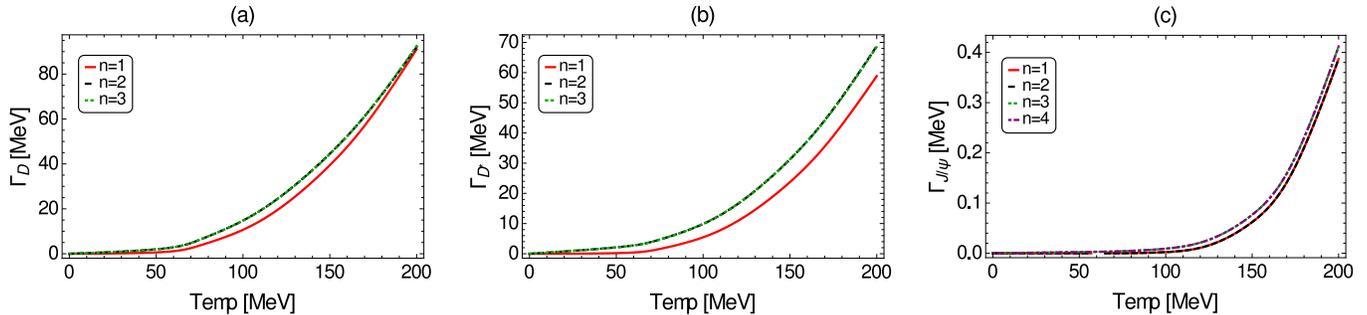}
\caption{
Pion-induced width as a function of the temperature at the meson mass and  $\vec{p} = 0$.  a): $D$, (b): $D^*$ and (c): $\jpsi$. To illustrate the process of achieving self consistent
results we show this for different iterations n. Further details are given in the text.
} \label{Fig:Width}
\end{figure*}

To demonstrate the impact of increasing temperatures we use a more illustrative quantity, the width of the mesons $M$ obtained from the surrounding pions
\begin{equation}
 \Gamma_M = \mathrm{Im}\Pi_M(p^0=m_M,\vec p=0)/m_M.
\end{equation}
Fig.~\ref{Fig:Width} shows the temperature behaviour of the three mesons in question, $D$, $D^*$ and $\jpsi$. 
Let us first discuss the final results which correspond to the last iteration at each sub-figure of Fig.~\ref{Fig:Width}, namely $n=3$ for the charmed meson case and $n=4$ for $\jpsi$. 
In the case of the charmed mesons we see a slow rise in the single $\mev$ region for temperatures below $100\mev$ and a sharper rise beyond that up to around $80-90\mev$ for temperatures approaching $200\mev$. 
The behaviour is similar to what He et al. see in~\cite{He:2011yi} although the values obtained in that work are smaller -- about 55~MeV at $T=200\mev$. However, given the differences in the approach -- unitarised effective field theory and resonance model, respectively -- the results are compatible with each other.
Compared to the hadronic width of the $D^*$ in the vacuum of the order of $100\kev$ this means an increase of 2-3 orders of magnitude depending on the temperature. 

% -------------------------------------------------------------
% -------------------------------------------------------------

The general behaviour is similar for the $\jpsi$ albeit suppressed by 2-3 orders of magnitude. 
At temperatures around $150\mev$ the pion induced width exceeds the vacuum width $\Gamma_{\jpsi}^\mathrm{Vac}=(93\pm3)\kev$~\cite{Olive:2016xmw}. 

As for the convergence of the iterative procedure, let us note that $n=1$ corresponds to the first iteration which employs free propagators for all mesons involved in the unitarized amplitude, see Figs.~\ref{Fig:diagram}(a). Iteration $n=2$ implements the dressing of the meson of which the width is being calculated. This is the reason why one observes a substantial change in going from $n=1$ to $n=2$ in the case of the $D$ and $D^*$ widths, which are now obtained with substantially modified spectral functions, while no visible change is seen in the $\jpsi$ width, which involves a very narrow, almost delta-like strength.  Convergence is achieved for $n=3$ in the charmed meson cases, as this iteration does not involve the dressing of any other type of mesons. However, iteration $n=3$ in the case of the  $\jpsi$ corresponds to additionally dressing the charmed mesons and, due to their broad spectral functions and to the fact that the unitarized amplitude for $\jpsi\pi$ scattering is driven by the $DD^*$ loop, a significant change in the $\jpsi$ width is observed with respect to iteration $n=2$. Convergence is reached for $n=4$ in this case.

%==================================================
\section{Summary}
%==================================================
We have calculated the behaviour of the charmed mesons and the $\jpsi$ in a pion bath at finite temperatures using unitarized $SU(4)$ chiral amplitudes. 
Compared to previous studies on the subject, this work is put on theoretically more sound foundations by using unitarized amplitudes and the Imaginary Time Formalism. 
Using this formalism we obtained observables that are relevant for understanding the behaviour of mesons containing charm quarks in hot mesonic matter, most notably the induced width and spectral function.

We find that the charmed $D$ and $D^*$ mesons acquire a substantial width, reaching values in the $30-40$ MeV range at $T=150$. Obviously, this will have consequences in the detected flux of these mesons in a hot pionic medium. Another interesting consequence of this finding refers to the properties of resonances  that have been explained in some models as molecules involving $D$ and/or $D^*$ mesons, as the $X(3872)$. How these resonances behave in a hot pionic medium is worth being explored in the light of the results presented here, a task that is presently under study.

The modification of the $\jpsi$ meson properties by a hot pion bath is substantially more moderate, due to the suppressed $\jpsi\pi\to\jpsi\pi$ interaction. 
We find that temperatures of  around $T=150$ MeV  are required for the the pionic width of the $\jpsi$ spectral function to exceed the free one. Therefore, the  
 $\jpsi$ retains most of its shape in hot pionic matter. Consequently, an observed drop in   $\jpsi$ detection should indeed point towards the formation of a QGP, the interaction with the surrounding hadrons being essentially negligible.
 
Our results can be used as input in theoretical simulations of open and hidden charm meson propagation in a hot pion gas, aiming at understanding the transition from a hadronic to a QGP phase.

\section{Acknowledgements}
This work is partly supported by the Spanish Ministerio de Economia y Competitividad (MINECO) under the project MDM-2014-0369 of ICCUB (Unidad de Excelencia 'Mar\'\i a de Maeztu'), 
and, with additional European FEDER funds, under the contract FIS2014-54762-P and the Spanish Excellence Network on Hadronic Physics FIS2014-57026-REDT.
Support has also been received from the Ge\-ne\-ra\-li\-tat de Catalunya contract 2014SGR-401.

%==================================================
\begin{appendix}
%==================================================
%--------------------------------------------------
\section{MESON-MESON STATES IN ISOSPIN BASIS}\label{app:amplitudes}
%--------------------------------------------------
Throughout this work we use Isospin eigenstates denoted as $|..>_I$. To calculate those we use the following phase conventions:
\begin{equation}
 \big|D^{(*)}\big>_{1/2} = \left( \begin{array}{c}   \phantom{-}\big|D^{(*)+}\big> \\ -\big|D^{(*)0}\big>      \end{array} \right)  \qquad
 \big|\pi\big>_{1} = \left( \begin{array}{c}   -\big|\pi^+\big> \\ \phantom{-}\big|\pi^0\big>  \\ \phantom{-}\big|\pi^-\big>  \end{array} \right)
\end{equation}
The relevant relations can then be derived as
\begin{equation}
 \left( \begin{array}{c}   \big|D^{(*)}\pi\big>_{1/2} \\ \big|D^{(*)}\pi\big>_{3/2}  \end{array} \right) = 
 \left( \begin{array}{cc}  -\sqrt{1/3} & -\sqrt{2/3} \\ -\sqrt{2/3} & +\sqrt{1/3} \end{array} \right)
 \left( \begin{array}{c}   \big|D^{(*)0}\pi^0\big> \\ \big|D^{(*)+}\pi^-\big>  \end{array} \right) \qquad
 \big|D^{(*)}\eta\big>_{1/2} = -\big|D^{(*)0}\eta\big>
\end{equation}
For the charmed meson pair with positive $C$-parity we find
\begin{equation}
 \left( \begin{array}{c} \big|D^{*}\bar D + D\bar D^*\big>_{0} \\ \big|D^{*}\bar D + D\bar D^*\big>_{1}  \end{array} \right) = 
 \left( \begin{array}{cc}  +1/2 & +1/2 \\ -1/2 & -1/2 \end{array} \right)
 \left( \begin{array}{c} \big|D^{*+} D^-\big> + \big|D^{+} D^{*-}\big> \\ \big|D^{*0} \bar D^0\big> + \big|D^{0} \bar D^{*0}\big> \end{array} \right)
\end{equation}

\begin{equation}
 \big|\jpsi\pi>_{1}  =  \big|\jpsi\pi^0>, \qquad
 \big|\eta_c\rho>_{1}  =  \big|\eta_c\rho^0>
\end{equation}
%==================================================
\end{appendix}
%==================================================

\end{document}